\documentclass[twocolumn,showpacs,prl,aps]{revtex4}
\usepackage{graphicx}
\usepackage{color}

\newcommand{\be}{\begin{equation}}
\newcommand{\ee}{\end{equation}}
\newcommand{\brma}{${\rm Ba_{0.4}Rb_{0.6}Mn_2As_2}$}

\begin{document}
\title{Ba$_{0.4}$Rb$_{0.6}$Mn$_2$As$_2$: A Prototype Half-Metallic Ferromagnet}
\author{Abhishek Pandey}
\altaffiliation{abhishek.phy@gmail.com}  
\author{D. C. Johnston}
\altaffiliation{johnston@ameslab.gov}
\affiliation {Ames Laboratory and Department of Physics and Astronomy, Iowa State University, Ames, Iowa 50011, USA}

\date{\today}

\begin{abstract}

Half-metallic ferromagnetism (FM) in single-crystal Ba$_{0.39(1)}$Rb$_{0.61(1)}$Mn$_2$As$_2$ below its Curie temperature $T_{\rm C} = 103(2)$~K is reported. The magnetization $M$ versus applied magnetic field $H$ isotherm data at 1.8~K show complete polarization of the itinerant doped-hole magnetic moments that are introduced by substituting Rb for Ba.  The material exhibits extremely soft FM, with unobservably small remanent magnetization and coercive field.  Surprisingly, and contrary to typical itinerant FMs, the $M(H)$ data follow the Arrott-plot paradigm that is based on a mean-field theory of local-moment FMs.  The in-plane electrical resistivity data are fitted well by an activated-$T^2$ expression for $T \leq T_{\rm C}$, whereas the data sharply deviate from this model for $T > T_{\rm C}$\@. Hence the activated-$T^2$ resistivity model is an excellent diagnostic for determining the onset of half-metallic FM in this compound, which in turn demonstrates the presence of a strong correlation between the electronic transport and magnetic properties of the material. Together with previous data on 40\% hole-doped Ba$_{0.6}$K$_{0.4}$Mn$_2$As$_2$, these measurements establish 61\%-doped Ba$_{0.39}$Rb$_{0.61}$Mn$_2$As$_2$ as a prototype for a new class of half-metallic ferromagnets in which all the itinerant carriers in the material are ferromagnetic. 

\end{abstract}


\maketitle

The discovery of high-$T_{\rm c}$ superconductivity in layered tetragonal LaFeAsO$_{1-x}$F$_x$ (1111-type) compounds in 2008 \cite{Kamihara-2008} fueled the search for superconducting phases in isostructural and also structurally-related FeAs-based materials. As a result, new 1111-type superconductors were discovered, together with other superconductors with related 11-, 111-, 122-type and other structures \cite{Johnston-2010, Paglione-2010,Stewart-2011}.  While most of the investigations in the field were centered around FeAs-based compounds, work was also carried out to extend the search for superconductivity and other exotic ground states to isostructural compounds where iron was competely replaced by another transition metal \cite{Singh-2009,An-2009,Pandey-2013a, Jayasekara-2013}.  BaMn$_2$As$_2$ is one such compound that has been extensively investigated \cite{Singh-2009,An-2009,Johnston-2011}. Special attention was paid to BaMn$_2$As$_2$ because it bridges the gap between the superconducting iron-arsenide and cuprate families of high-$T_{\rm c}$ compounds. On the one hand it crystallizes in the ThCr$_2$Si$_2$-type body-centered tetragonal structure of the model BaFe$_2$As$_2$ parent compound, while on the other hand it has an insulating antiferromagnetic (AFM) ground state and contains a stacked square lattice of local Mn moments, properties that are similar to those of the cuprate superconductor parent compounds containing Cu local moments \cite{Johnston-1997}.

Although BaMn$_2$As$_2$ is isostructural to BaFe$_2$As$_2$, the properties of the two compounds are highly divergent \cite{Johnston-2010, Singh-2009, An-2009}.  The insulator BaMn$_2$As$_2$ shows G-type AFM ordering of local Mn moments with a N\'eel temperature $T_{\rm N} = 625$~K \cite{Singh-2009}. Interestingly, a metallic ground state can be induced by substituting a  small amount of K for Ba \cite{Pandey-2012,Bao-2012,Yeninas-2013} or by applying pressure \cite{Satya-2011}. Furthermore, ferromagnetism (FM) occurs in hole-doped single crystals of Ba$_{1-x}$K$_x{\rm Mn_2As_2}$ with $x = 0.19$ and~0.26 with saturation (ordered) moments of 0.01 and $0.15~\mu_{\rm B}$/f.u., respectively \cite{Bao-2012}, where $\mu_{\rm B}$ is the Bohr magneton and f.u.\ stands for formula unit.  A unique magnetic ground state occurs in the more highly hole-doped Ba$_{0.60}$K$_{0.40}$Mn$_2$As$_2$ where itinerant FM arising from {\it complete} polarization of the doped conduction holes [half-metallic (HM) ferromagnetism] \cite{deGroot-1983, Park-1998, Graf-2011} below the Curie Temperature $T_{\rm C} \approx 100$~K coexists with robust collinear local moment AFM of Mn spins with $T_{\rm N} = 480$~K  \cite{Pandey-2013b, Lamsal-2013}. Remarkably, the easy axes of the respective collinear FM  and AFM  ordered moments are orthogonal to each other, with the AFM Mn moments aligned along the $c$~axis and the doped-hole FM moments aligned in the $ab$~plane.  Recently, element-specific x-ray magnetic circular dichroism measurements demonstrated that the FM indeed resides on the As $4p$ orbitals with no observable contribution from the Mn $3d$ orbitals \cite{Ueland-2015}, which disproves a suggestion that the FM originates from canting of the Mn ordered moments towards the $ab$~plane \cite{Glasbrenner-2014}.  In general it is difficult to unambiguously establish whether or not a given itinerant FM compound is a half-metallic FM \cite{Park-1998, Hanssen-1990, Coey-2002, Katsnelson-2008}.  However, in Ba$_{0.60}$K$_{0.40}$Mn$_2$As$_2$ the HM nature of the FM state at low~$T$ is obvious because all the itinerant doped-hole magnetic moments are found to be polarized in the FM state. To our knowledge, Ba$_{0.60}$K$_{0.40}$Mn$_2$As$_2$ is the first ThCr$_2$Si$_2$-type compound that shows HM-FM behavior.

Here we significantly extend the previous work on HM-FM in heavily hole-doped Ba$_{1-x}A_{x}$Mn$_2$As$_2$ ($A$: Alkali metal) materials.  We report the growth of Rb-doped Ba$_{0.39(1)}$Rb$_{0.61(1)}$Mn$_2$As$_2$ crystals (abbreviated as \brma) with an ordered moment of $0.64(1)~\mu_{\rm B}$/f.u.\ at temperature $T = 1.8$~K that is about 50\% larger than in Ba$_{0.60}$K$_{0.40}$Mn$_2$As$_2$, consistent with the above interpretation of the FM as arising from complete polarization of the doped-hole moments.  The Curie temperature $T_{\rm C} = 103(2)$~K is about the same as in Ba$_{0.60}$K$_{0.40}$Mn$_2$As$_2$.  Surprisingly, $ab$-plane magnetization~$M$ versus applied magnetic field~$H$ isotherms at temperatures spanning $T_{\rm C}$ plotted as $M^2$ versus $H/M$ follow Arrott's mean-field theory prediction  for local-moment FMs \cite{Arrott-1957, Arrott-2010}, and the isotherm at 1.8~K exhibits an unobservably small remanent magnetization and coercive field.   As in Ba$_{0.60}$K$_{0.40}$Mn$_2$As$_2$ \cite{Pandey-2013b}, the in-plane electrical resisitivity $\rho_{ab}(T)$ data for \brma\ at $T < T_{\rm C}$ agree with the phenomenological equation \cite{Bombor-2013} that describes thermally-activated carrier scattering between the spin-split bands in a HM-FM, whereas the model strongly deviates from the data for \brma\ at $T>T_{\rm C}$, thus providing an independent determination of $T_{\rm C}$.  These measurements establish \brma\ as a prototype of a new class \cite{Coey-2002} of HM-FMs in which all of the itinerant carriers in the material are ferromagnetic.

Single crystals of Ba$_{0.4}$Rb$_{0.6}$Mn$_2$As$_2$ were grown using a MnAs self-flux solution-growth technique similar to that described in \cite{Lamsal-2013}. Shiny plate-like crystals with typical dimensions $5 \times 5 \times 0.3$~mm$^3$ were obtained that are quite stable in air.  The compositions of the crystals were determined by energy- and wavelengh-dispersive x-ray spectroscopy measurements. A crystal with composition Ba$_{0.39(1)}$Rb$_{0.61(1)}$Mn$_2$As$_2$, which is referred to as Ba$_{0.4}$Rb$_{0.6}$Mn$_2$As$_2$ in the rest of the paper, was selected for the measurements.  Magnetization and four-probe $\rho_{ab}(T)$ measurements were carried out using Quantum Design, Inc., MPMS and PPMS instruments, respectively.  Room-temperature powder x-ray diffraction (XRD) measurements on a representative crushed crystal were carried out using a Rigaku Geigerflex powder diffractometer employing Cu-$K_{\alpha}$ radiation. The powder XRD data were refined using the {\tt FullProf} Rietveld refinement package \cite{Carvajal-1993} and the refined values of the tetragonal lattice parameters are $a = 4.1846(2)$~\AA\ and $c = 13.3920(9)$~\AA\@.  The $a$ lattice parameter is larger than, and the $c$ lattice parameter smaller than, the respective values $a = 4.1674(6)$~\AA\ and $c = 13.467(2)$~\AA\ obtained for the parent BaMn$_2$As$_2$ compound \cite{Singh-2009}.

\begin{figure}
\includegraphics[width=3.3in]{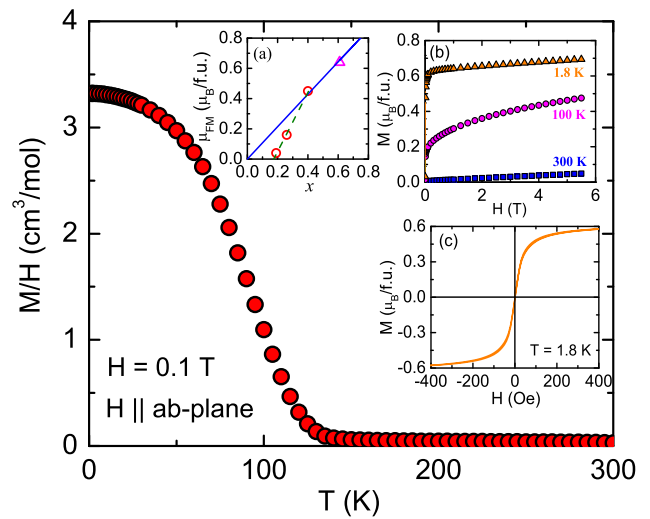}
\caption{(Color online) Magnetization $M$ divided by applied magnetic field $H$ versus temperature $T$ data for a crystal of Ba$_{0.4}$Rb$_{0.6}$Mn$_2$As$_2$ with $H$ applied parallel to the $ab$~plane. Inset (a): Variation of the FM ordered moment $\mu_{\rm FM}$ versus the concentration~$x$ of doped holes from K (open red circles) and Rb (open magenta triangle) doping.	The data for $x = 0.19$ and 0.26 are from \cite{Bao-2012} and the datum for $x = 0.40$ is from \cite{Pandey-2013b}. The solid blue line is a proportional fit according to $\mu_{\rm FM}(\mu_{\rm B}/{\rm f.u.}) = xgS\mu_{\rm B}$ ($S = 1/2$ for the hole spins and $g = 2.14$) to the two data points $x = 0.40$ and $0.61$. The dashed green line is a guide to eye. Inset~(b): Isothermal $M$ versus $H$ measured at three different temperatures. Inset (c): Low field $M(H)$ hysteresis loop measured at 1.8~K.}
\label{fig:Figure_Mag}
\end{figure}

The in-plane magnetic susceptibility $\chi_{ab} \equiv M/H$ versus $T$ data at $H=0.1$~T, shown in Fig.~\ref{fig:Figure_Mag}, exhibit a huge FM enhancement below $\sim 125$~K\@. This behavior is attributed to a strong increase in the  polarization of the doped-hole magnetic moments leading to HM behavior at low temperatures \cite{Pandey-2012}. It is evident from the three representative $M(H)$ plots shown in inset~(b) of Fig.~\ref{fig:Figure_Mag} that a saturation (ordered) moment $\mu_{\rm FM}$ develops at $T \lesssim 100$~K\@. By extrapolating the high-field $M(H)$ data at 1.8~K to $H = 0$, we obtain $\mu_{\rm FM} = 0.64(1)~\mu_{\rm B}$/f.u. Taking the spectroscopic splitting factor to be $g = 2$, the ordered moment for completely polarized doped-hole spins $S = 1/2$, given by
\be
\mu_{\rm FM} = xgS\mu_{\rm B}/{\rm f.u.},
\label{Eq:muFM}
\ee
would be $\mu_{\rm FM} = 0.61~\mu_{\rm B}$/f.u.\ for a Rb content $x=0.61$.  This value is in close agreement with the above value of $\mu_{\rm FM}$ observed at 1.8~K, demonstrating that the doped holes are completely spin polarized at this temperature.

Inset~(a) of Fig.~\ref{fig:Figure_Mag} shows $\mu_{\rm FM}$ versus the doped-hole concentration~$x$. Fitting the data points for the two compositions $x = 0.40$ and $0.61$ that show HM-FM behavior by Eq.~(\ref{Eq:muFM}) assuming $S=1/2$ gives $g = 2.14$ which is close to the value $g=2$ for free $S=1/2$ conduction holes.  Low-field hysteresis-loop data taken at 1.8~K $\ll T_{\rm C}$ such as shown in inset~(c) of Fig.~\ref{fig:Figure_Mag} exhibit extremely soft FM behavior with very small values of coercive field and remanant magnetization, as observed for other HM-FMs \cite{Coey-1998, Ritchie-2003}.  

\begin{figure}
\includegraphics[width=3.3in]{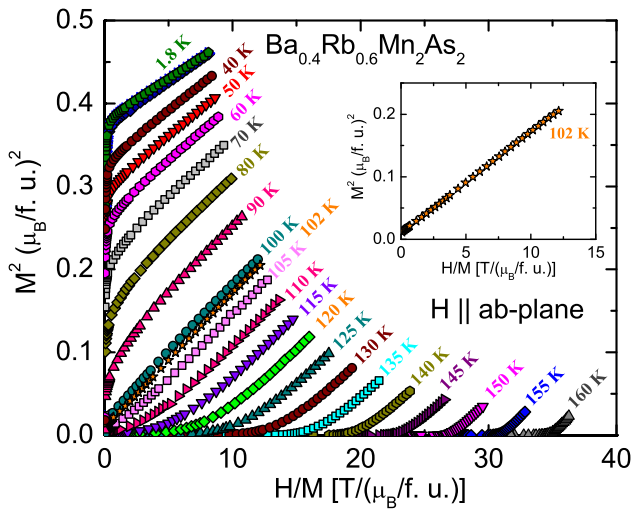}
\caption{(Color online) Arrott plot: squared in-plane magnetization $M^2$ versus applied magnetic field $H$ divided by $M$. Inset: Expanded plot of $M^2$ versus $H/M$ data taken at $T = 102$~K\@.}
\label{fig:Figure_Arrot-1}
\end{figure}

\begin{figure}
\includegraphics[width=3.3in]{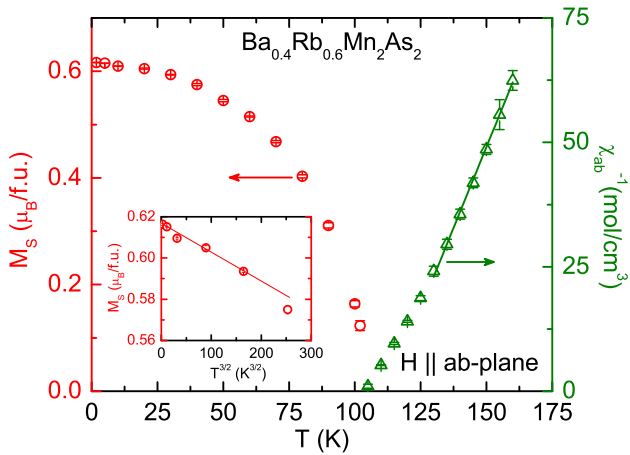}
\caption{(Color online) Temperature $T$ variations of the saturation magnetization $M_{\rm s}$ and inverse molar susceptibility ($\chi_{ab}^{-1} = H/M$) extracted from the Arrott plot measurements in Fig.~\ref{fig:Figure_Arrot-1} are plotted with left and right ordinate scales, respectively. The solid green straight line is the fit described in the text. Inset: Variation of $M_{\rm s}$ with $T^{3/2}$. The solid red straight line is a guide to eye.}
\label{fig:Figure_Arrot-2}
\end{figure}

Figure~\ref{fig:Figure_Arrot-1} presents $M(H)$ data at twenty-one temperatures between 1.8 and 160~K, plotted as $M^2$ versus $H/M$.  These Arrott-plot data show the behavior expected for local-moment FMs, contrary to the typical behavior  for weak ferromagnets where $M^4$ versus $H/M$ plots are more appropriate \cite{Takahashi-2012, Sato-2010, Ghimire-2012}. The reason that our $M^2$ versus $H/M$ data follow the mean-field  prediction for local moments is not clear and deserves theoretical investigation.  The data taken at 102~K show linear and nearly proportional behavior (inset, Fig.~\ref{fig:Figure_Arrot-1}), leading to a FM transition temperature $T_{\rm C} = 103(2)$~K\@. The values of $M_{\rm s}$ (i.e., $\mu_{\rm FM}$) and inverse molar susceptibility $\chi_{ab}^{-1}$ deduced from the extrapolated intercepts of the the high-field fits to the $y$- and $x$-axes, respectively, are shown in Fig.~\ref{fig:Figure_Arrot-2}. The $M_{\rm s}(T)$ data show a behavior that is qualitatively similar to those of most magnetically-ordered materials.  At low-$T$ $M_{\rm s}$ varies as $T^{3/2}$ (inset, Fig.~\ref{fig:Figure_Arrot-2}), a behavior expected at low temperatures ($\lesssim T_{\rm C}/3$) from FM spin-wave excitations \cite{Makoshi-1975, Hordequin-1996}. The $\chi_{ab}^{-1}(T)$ shows nonlinear behavior close to $T_{\rm C}$, presumably due to FM correlations not included in mean-field theory; however, we do observe a nearly linear region between 130 and 160~K\@.  A linear fit to the $\chi_{ab}^{-1}(T)$ data over this $T$~range yields a Curie constant $C = 0.78$~cm$^3$~K/mol. This $C$ is larger than the value $C = \frac{g^2}{8}S(S+1) = 0.43$~cm$^3$~K/mol expected for a spin-1/2 local-moment system with the above $g = 2.14$. Possible reasons for this discrepancy might be the proximity of the fitted data points to $T_{\rm C}$, the itinerant character of the doped-hole spins rendering the local-moment picture inapplicable, and/or a $T$-dependent background associated with the localized Mn spins that are AFM-ordered in the $T$-range of this fit.   Figure~5 of~\cite{Lamsal-2013} suggests that $T_{\rm N}$ of the Mn sublattice for $x = 0.6$ should be roughly 300~K\@.

\begin{figure}
\includegraphics[width=3.3in]{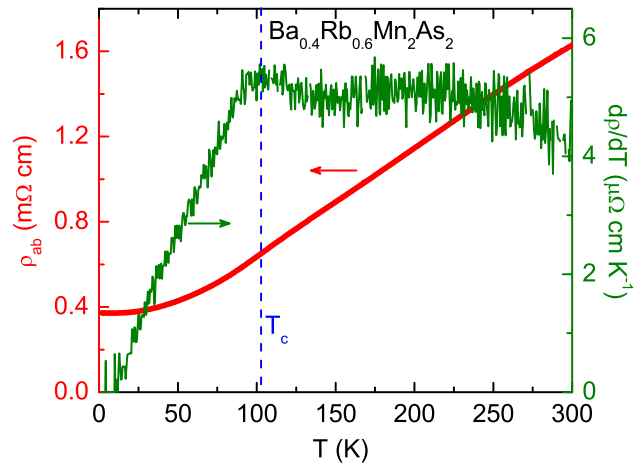}
\caption{(Color online) Left ordinate: In-plane electrical resistvity $\rho_{ab}$ versus temperature $T$. Right ordinate: Temperature derivative of $\rho_{ab}$ versus $T$. The vertical dashed blue line shows the ferromagnetic transition temperature $T_{\rm C} = 103$~K obtained from the Arrott plot measurements.}
\label{fig:Figure_Res1}
\end{figure}

The $\rho_{ab}(T)$ along with its first $T$-derivative are shown in Fig.~\ref{fig:Figure_Res1}.  While $\rho_{ab}(T)$ shows a metallic (positive) temperature coefficient of resistivity, the magnitude of $\rho_{ab}$ is large for a metal.  As evident from the $d\rho/dT$ versus $T$ plot in Fig.~\ref{fig:Figure_Res1}, the slope exhibits a rapid increase with increasing $T$ up to $T_{\rm C}$ and then becomes nearly $T$-independent for $T > T_{\rm C}$\@. The plateau in the $d\rho/dT$ versus $T$ plot may be due to a decrease of spin-disorder scattering above $T_{\rm C}$. Overall, the $\rho_{ab}(T)$ data in Fig.~\ref{fig:Figure_Res1} exhibit an interesting correlation between the FM order parameter and the electronic transport properties in \brma, as expected since both the FM and electronic transport are associated with the same conduction carriers.

\begin{figure}
\includegraphics[width=3.3in]{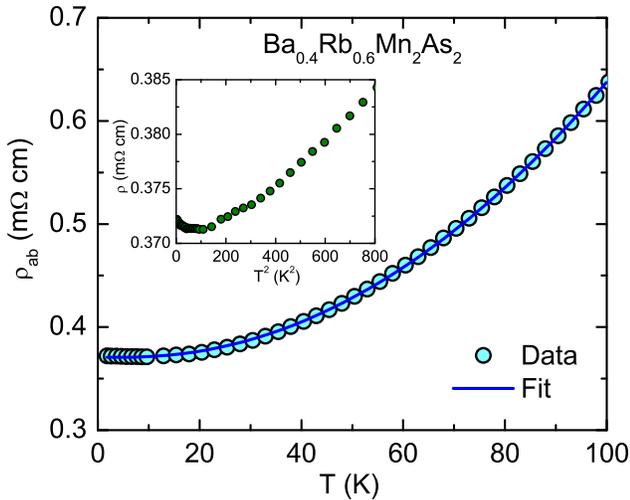}
\caption{(Color online) In-plane resistivity $\rho_{ab}$ versus temperature $T$ between 1.8 and 103~K\@. The solid curve is a fit of the data by Eq.~(\ref{Eq:rhoFit}). Inset: $\rho_{ab}$ versus $T^2$ at low temperatures.  This plot shows that $\rho_{ab}$ is not proportional to $T^2$ below 28~K\@.}
\label{fig:Figure_Res2}
\end{figure}

\begin{figure}
\includegraphics[width=3.3in]{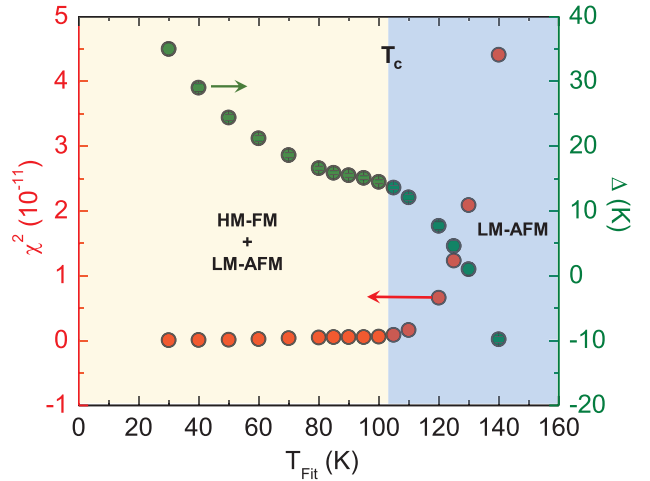}
\caption{(Color online) Dependence on $T_{\rm Fit}$ of the goodness of fit parameter $\chi^2$ (left ordinate) and the energy gap $\Delta$ of the spin-split band (right ordinate) obtained by fitting the $\rho_{ab}(T)$ data for \brma\ between 1.8~K and a temperature $T_{\rm Fit}$ by Eq.~(\ref{Eq:rhoFit}).  The yellow area to the left of $T_{\rm C}=103$~K is the temperature range of coexisting itinerant half-metallic ferromagnetism (HF-FM) and Mn local-moment antiferromagnetism (LM-AFM) and the blue area to the right of $T_{\rm C}$ corresponds to the region where only the LM-AFM occurs.  The strong worsening of $\chi^2$ with increasing $T_{\rm Fit}$ occurs as $T_{\rm Fit}$ traverses $T_{\rm C}$ in increasing $T_{\rm Fit}$.}
\label{fig:Figure_Phase}
\end{figure}

Single-magnon scattering that involves a spin flip process results in a $T^2$ behavior of $\rho$ at low temperatures in FM materials \cite{Mannari-1959}. Because of the completely polarized character of the bands at the Fermi level, this scattering is absent at $T \to 0$ in ideal HM-FMs.  However, as $T$ increases, majority-spins are thermally excited to the minority-spin band and a $T$-assisted scattering results with a $\rho(T)$ given by \cite{Bombor-2013}
\be
\rho(T) = \rho_{0} + AT^{2}e^{-\Delta/T},
\label{Eq:rhoFit}
\ee
where $\Delta$ is the energy gap between the Fermi level of the material and bottom of the minority spin conduction band. Our $\rho_{ab}(T)$ data do not show a $T^2$ dependence at low temperatures (inset, Fig.~\ref{fig:Figure_Res2}).  Instead an excellent fit of Eq.~(\ref{Eq:rhoFit}) to the data is obtained for $T < T_{\rm C}$ (Fig.~\ref{fig:Figure_Res2}), as was also obtained for $\rho(T)$ of Ba$_{0.60}$K$_{0.40}$Mn$_2$As$_2$ \cite{Pandey-2013b}.  However the fit quality degrades severely when the maximum temperature of the fit exceeds $T_{\rm C}$, as shown in Fig.~\ref{fig:Figure_Phase}.  We did not need to include an electron-phonon scattering contribution to $\rho_{ab}(T)$ in order to obtain the excellent fit to our data below $T_{\rm C}$.  This observation and the large magnitude of $\rho_{ab}(T)$ suggest that below $T_{\rm C}$ the electron-magnon scattering rate is large compared to the electron-phonon scattering rate.

The band gap $\Delta$ shows a slow decrease followed by a shallow plateau with increasing $T_{\rm Fit}$ and then undergoes a clear change of slope at $T_{\rm C}$ above which it decreases strongly and attains unphysical negative values for $T_{\rm Fit} > 130$~K (Fig.~\ref{fig:Figure_Phase}). The value of $\Delta$ obtained at low temperatures is comparable to those reported for other HM-FMs \cite{Bombor-2013}. The goodness-of-fit parameter $\chi^{2}$ shows small and nearly $T$-independent values for $T_{\rm Fit} \leq T_{\rm C}$, where the data have been fitted between the lowest temperature of measurement 1.8~K and a variable temperature $T_{\rm Fit}$. However, $\chi^{2}$ increases rapidly for $T_{\rm Fit} > T_{\rm C}$, exhibiting a strong deviation from the activated-$T^2$ dependence in Eq.~(\ref{Eq:rhoFit}). This again confirms the presence of a strong correlation between the FM and the electronic transport properties of this system. 

In conclusion, we grew crystals of the heavily hole-doped tetragonal ThCr$_2$Si$_2$-type compound Ba$_{0.39(1)}$Rb$_{0.61(1)}$Mn$_2$As$_2$ and disovered HM-FM ordering in this material  below $T_{\rm C} = 103(2)$~K\@. The compound exhibits extremely soft FM with very small (i.e., unobservably small) coercive field and remanent magnetization. The value of the saturation moment $\mu_{\rm FM} = 0.64(1)~\mu_{\rm B}$/f.u.\ at low temperatures indicates complete polarization of the 0.61(1) doped-hole magnetic moments and is a prototype of a new type of HM-FM \cite{Coey-2002} in which all the itinerant current carriers in the material are spin polarized.  Ba$_{0.39}$Rb$_{0.61}$Mn$_2$As$_2$ shows highly correlated FM and electronic transport properties as expected for a HM-FM system but, surprisingly, exhibits Arrott-plot-type mean-field $M(H)$ behavior characteristic of local-moment systems.  The previous data on ThCr$_2$Si$_2$-type Ba$_{0.60}$K$_{0.40}$Mn$_2$As$_2$ \cite{Pandey-2013b,Lamsal-2013,Ueland-2015} and especially our new data on Ba$_{0.39}$Rb$_{0.61}$Mn$_2$As$_2$ suggest many opportunities for further experimental and theoretical studies of the interesting features of this new class of HM-FMs.

\acknowledgments

The work at Ames Laboratory was supported by the U.S.~Department of Energy, Office of Basic Energy Sciences, Division of Materials Sciences and Engineering.  Ames Laboratory is operated for the U.S.~Department of Energy by Iowa State University under Contract No.~DE-AC02-07CH11358.

\end{document}